\documentclass[10pt,twocolumn,letterpaper]{article}

\usepackage{cvpr}      

\usepackage{times}
\usepackage{helvet}
\usepackage{courier}
\usepackage{xcolor}
\usepackage{pifont}

\usepackage{booktabs}
\usepackage{multirow}
\usepackage{graphicx}   
\usepackage{subcaption}  
\definecolor{cvprblue}{rgb}{0.21,0.49,0.74}
\usepackage[pagebackref,breaklinks,colorlinks,allcolors=cvprblue]{hyperref}

\def\paperID{} 
\def\confName{CVPR}
\def\confYear{2026}

\title{SiGRRW: A Single-Watermark Robust Reversible Watermarking Framework with Guiding Strategy}

\author{Zikai Xu, Bin Liu, Weihai Li, Lijunxian Zhang, Nenghai Yu\\
{\tt \{ustcxzk@mail., flowice@, whli@, ljxzhang@mail., ynh@\}ustc.edu.cn}\\
University of Science and Technology of China\\
Anhui, China
}

\begin{document}
\maketitle
\begin{abstract}
Robust reversible watermarking (RRW) enables copyright protection for images while overcoming the limitation of distortion introduced by watermark itself. Current RRW schemes typically employ a two-stage framework, which fails to achieve simultaneous robustness and reversibility within a single watermarking, and functional interference between the two watermarks results in performance degradation in multiple terms such as capacity and imperceptibility.
We propose SiGRRW, a single-watermark RRW framework, which is applicable to both generative models and natural images. We introduce a novel guiding strategy to generate guiding images, serving as the guidance for embedding and recovery. The watermark is reversibly embedded with the guiding residual, which can be calculated from both cover images and watermark images.  The proposed framework can be deployed either as a plug-and-play watermarking layer at the output stage of generative models, or directly applied to natural images. 
Extensive experiments demonstrate that SiGRRW effectively enhances imperceptibility and robustness compared to existing RRW schemes while maintaining lossless recovery of cover images, with significantly higher capacity than conventional schemes.
\end{abstract}    
\section{Introduction}
\begin{figure}[t]
	\centering
	\includegraphics[width=0.4\textwidth]{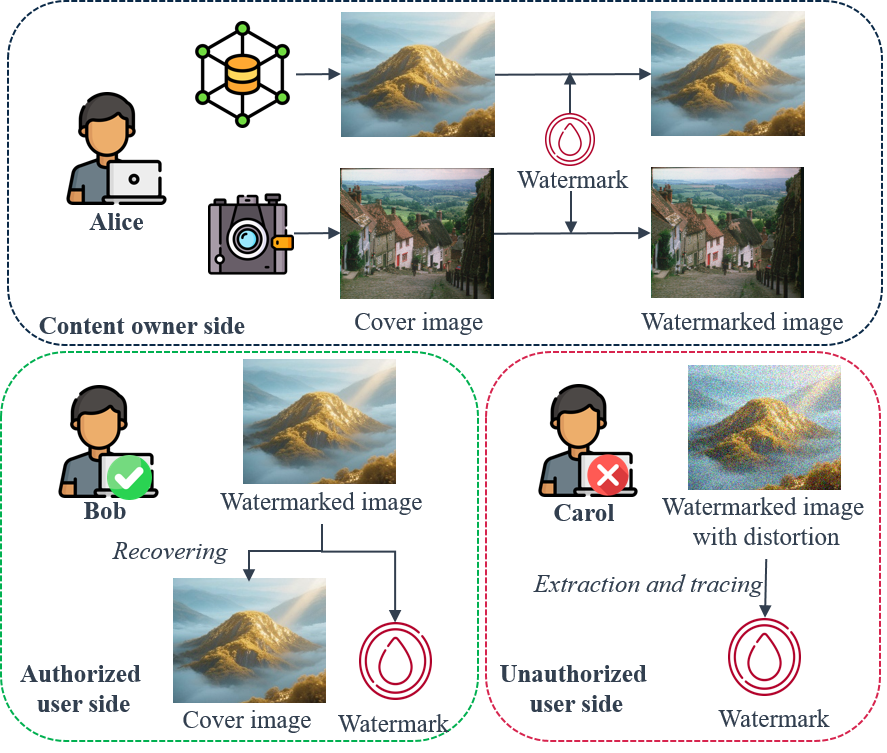} 
	\caption{An application scenario of RRW}
	\label{ApplicationScenario}
	
\end{figure}

Images serve as a fundamental component of modern communication, driving exponential transmission across network platforms. However, the transmission of natural images and model-generated images via social media are faced with deliberate or unconscious copyright infringement risks, including piracy and surrogate model attacks \cite{Zhang2022Deep, LiuCQ00P25}. Robust watermark is an effective copyright protection technique with resistance to distortions during transmission \cite{Wan2022, XiaoZHXW24, GuoLHGZCWW24}. And when images remain undistorted by external factors, distortions introduced by watermark itself should also be considered. Any distortion of original images may compromise specific applications such as medical imaging, professional photography and deep model training \cite{Wang2024}. Therefore, transmitted images should be losslessly recoverable by authorized users, and Robust Reversible Watermarking (RRW) has been developed to address both external and internal distortion issues \cite{Tang2023A}.
One application scenario of RRW is shown in \cref{ApplicationScenario}, where the content owner embeds watermark into the cover images generated by models or cameras, and then publishes them via the Internet. Unauthorized users may access and utilize these images, while watermark can be extracted from the images for traceability even after distortion. Authorized users can remove the watermarks to recover the original images without degradation under non-attack conditions, enabling further precision applications such as medical imaging research or model training.

Current robust reversible watermarking schemes usually employ a two-stage framework \cite{Wang2024}, achieving notable performance in both robustness and reversibility, and they outperform earlier frequency-domain feature computation approaches in various terms. The two-stage RRW is implemented by first embedding messages via a robust watermark, and then embedding a reversible watermark to carry auxiliary information for recovery of the robust embedding. Fundamentally, two-stage framework merely combines two distinct watermarks, and it does not constitute a genuine RRW scheme implemented with a single watermark. Compared with single-watermark schemes, such a combination inevitably degrades robustness, reduces embedding capacity, and increases computational cost. 
The reason is that the embedding of additional auxiliary data diminishes the overall capacity available for the intended message, and when suffering attacks, the damaged reversible watermark usually brings additional distortion into the robust one which further weakens its robustness. Moreover, existing reversible watermarking schemes mostly rely on conventional mathematical computations, which further constrain embedding capacity and imperceptibility \cite{Guo2025RRW}. 

To address the aforementioned constraints, we propose SiGRRW, a novel \textbf{Si}ngle-watermark \textbf{G}uided \textbf{R}eversible \textbf{R}obust \textbf{W}atermarking scheme for both natural images and model-generated images, performing robust embedding and lossless recovery.
We note that cover image $I_o$ and watermarked image $I_w$ maintain highly similar structures to preserve both visual and statistical consistency, thus identical features between them can be retrieved and utilized to guide reversible embedding. If the watermark is embedded following the guidance of identical feature representations,
the resulting embedding residual becomes uniquely determined by either $I_o$ or $I_w$, and can therefore be reliably computed from it. Then original image is recovered by removing the residual from $I_w$. 
We introduce a novel \emph{guiding strategy} to ensure reversibility, where \textbf{Guider} is utilized to output identical guiding images $I_g$ guaranteed by the similar net structure to the embedding subnet. The watermark is embedded into $I_o$ based on the trained guiding residual, which is generated from the embedding to $I_g$. Later the identical guiding image and residual can be acquired by \textbf{Guider} with $I_w$, and the original image can be recovered losslessly by removing guiding residual from $I_w$. Furthermore, the incorporation of a noise layer during training facilitates enhanced robustness while maintaining reversibility. 

After training, SiGRRW is free of model parameter modifications to be plug-and-play for models, and it also maintains compatibility with natural images. The main contributions are summarized as follows:
\begin{itemize}
	\item We propose a deep-model-based robust reversible watermarking framework, and with a guiding strategy, the cover image can be perfectly recovered from the watermarked image while maintaining robustness.  
	\item Component \textbf{Guider} is introduced to precisely generate identical guiding images from the cover image and the watermarked image. And the similar structures between the guiding subnet and the embedding subnet ensure maximal guiding consistency, while a novel adjustment method guarantees completely identical consistency.
	\item We introduce a simplified noise layer, using parallel limited noise branches to enhance robustness of the watermark against various distortions. 
	\item Extensive experiments illustrate that SiGRRW framework achieves lossless reversible recovery, large embedding capacity, high imperceptibility and robustness.
	
\end{itemize}

\section{Related Work}
\subsection{Model Watermarking}
Model watermarking can be categorized into two groups based on embedding methodology: pretrained watermarking and plug-and-play watermarking \cite{PlugandPlay}. Pretrained ones embed watermarks during generative model training by correlating them with model parameters, while plug-and-play ones are train-free with more flexibility and lower cost. 

Early pretrained-model watermarking methods primarily embed information through the training set \cite{Yu2021Artificial}, but they require retraining for each new watermark. To address this, Yu \cite{Yu2022Responsible} later proposed integrating watermarking into the generative process. Wu \cite{Wu2021watermarking} extended watermarking across various image processing tasks. Lin \cite{Lin2025CycleGAN} proposed a CycleGAN-based scheme that jointly trains the generator with a frozen watermark decoder.
With the rapid advancement of diffusion models, diffusion-based watermark has received increasing research attention \cite{Fernandez2023Stable, YangXLDQ25, Fang2025CoSDA, Yang2024Saussian, WangGZ0H0T25, LeiG0Z025, WenKGG23}. 
Yang \cite{Yang2024Saussian} proposes Gaussian Shading, proving it to be performance-lossless in diffusion model. 

Plug-and-play ones append an additional layer after image generation, offering reduced training costs and enhanced portability across different model.
Kou \cite{Kou2025Robust} proposes a Learnable Wavelet Network to balance imperceptibility and robustness.
Wang \cite{Wang2024Must} extends the traceability to multi-source images and minimizes the external rectangle operation.
Zhang \cite{Zhang2022Deep} proposes a deep model watermarking by adding a special task-agnostic barrier after the target model to embed a unified watermark into the output.
Zhang \cite{Zhang2024Robust} proposes a robust model watermarking based on structure consistency, further extending robustness.

\subsection{Robust Reversible Watermarking}
Typically, traditional reversible watermarking schemes are mainly achieved with mathematical techniques in spatial domain, such as difference expansion \cite{Tian2003}, histogram shifting \cite{Ni2006}, and prediction-error expansion \cite{Sachnev2009}. Most of these schemes are incapable under distorted conditions, and RRW schemes have been developed. Early RRW schemes typically extended reversible watermarking technique from spatial domain to robust features in frequency domain \cite{An2012Robust}. While these schemes achieved robustness against specific attacks, they often struggle with scalability for complex distortions \cite{Chen2025deep}.

In recent years, two-stage RRW framework has gained widespread adoption and research attention. Wang \cite{Wang2020} proposes a two-stage RRW framework based on independent embedding domains. 
Xiong \cite{Xiong2022Robust} introduced a ciphertext-image RRW scheme enhancing robustness against JPEG compression and noise attacks.
Tang \cite{Tang2023A} proposes a Zernike-Moment-based RRW scheme, utilizing embedding optimization and rounded error compensation to improve the imperceptibility of the watermark.
Wang \cite{Wang2024} extends the robustness of the two-stage RRW to cropping by embedding robust watermark into multiple local Zernike moments.
Chen \cite{Chen2025deep} developed an INN-based two-stage RRW framework that incorporates arithmetic coding and an END-based strategy for adaptive robustness under diverse distortions.
Guo \cite{Guo2025RRW} proposed the first two-stage RRW against regeneration attacks \cite{Zhao2024regeneration} based on orthogonal moment.

\begin{figure*}[t]
	\centering
	\includegraphics[width=\textwidth]{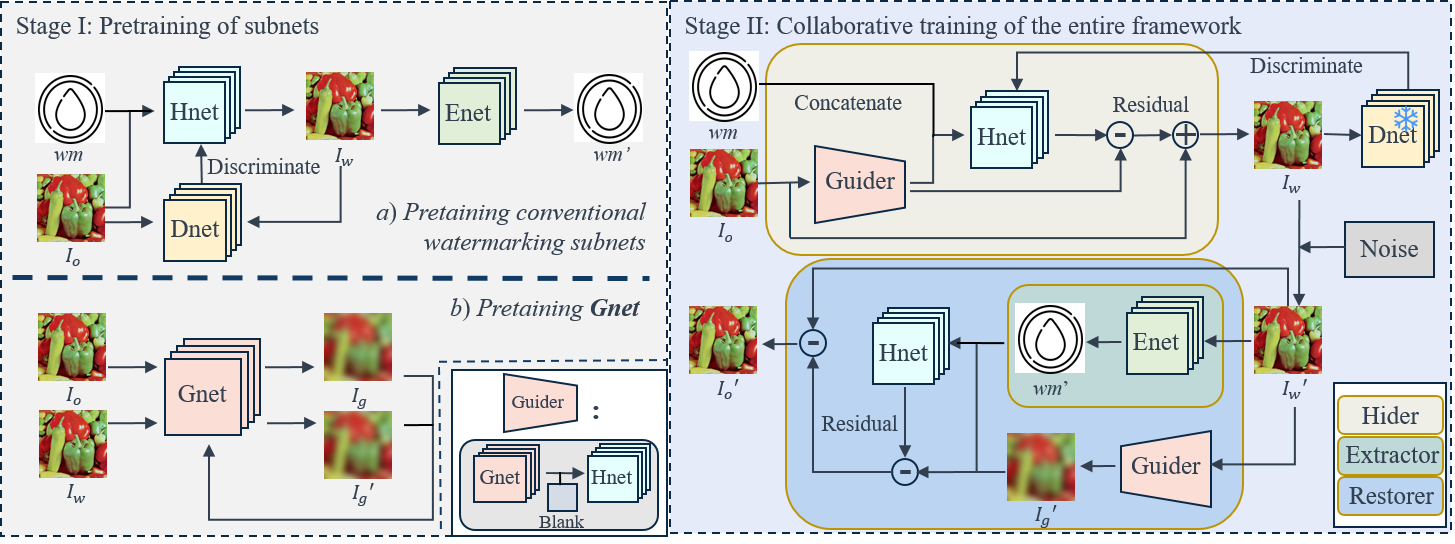} 
	\caption{The overall architecture and training procedure of the proposed framework}
	\label{flowchart}
\end{figure*}

\section{Threat Model}
We define the threat model considered for SiGRRW, which is commonly used in RRW methods. As shown in \cref{ApplicationScenario}, the scenario involves three parties: the content owner Alice, authorized user Bob, and unauthorized user Carol. 

Alice owns the copyright of images. She shares high-fidelity images over the Internet and employs watermarking to protect copyright. On one hand, she expects that unauthorized users cannot remove the watermark with common distortions to evade traceability; on the other hand, she desires that authorized users can utilize the images normally without being affected by the watermark-induced distortion.

Bob, who is authorized by Alice, can freely use the shared images. By employing \texttt{Restorer} provided by Alice, he is able to remove the watermark and perfectly reconstruct the original, watermark-free images losslessly ($\mathrm{PSNR} = \infty$).

Carol, an unauthorized user, attempts to exploit Alice’s images without permission and evade traceability. Without access to \texttt{Restorer}, she tries to remove the watermark through common distortions such as common signal processing (CSP) or JPEG compression, or even employs regeneration-based approaches \cite{Zhao2024regeneration} to erase the watermark. She then republishes the processed images for profit.

\section{Method}
\subsection{Overall Architecture}
The overall architecture and training procedure of SiGRRW framework are shown in \cref{flowchart}, which contains three modules with four subnets. The training of SiGRRW contains two stages, the pretraining of subnets and the collaborative training of the entire framework. After the collaborative training, three modules in Stage II pipeline require deployment in practical applications, namely the \texttt{Hider}, \texttt{Extractor} and \texttt{Restorer}, while \textbf{Dnet} serves solely as an auxiliary training module. \texttt{Hider} performs reversible embedding of watermark $wm$ into the cover image $I_o$.
\texttt{Extractor} extracts watermark $wm'$ from watermarked image $I_w$.
\texttt{Restorer} recovers original image $I_o'$ from $I_w$ and $wm'$ under the guiding strategy, and is distributed to authorized users. 

The subnets of conventional watermark embedding are inspired by \cite{Zhang2022Deep}. \textbf{H}($\cdot$), \textbf{E}($\cdot$), \textbf{D}($\cdot$) and \textbf{G}($\cdot$) denote the output of the subnets respectively in the following pages.
We deploy a widely used UNet \cite{unet} structure as \textbf{Hnet} and \textbf{Gnet}. We apply CEILNet \cite{CEILnet} structure for \textbf{Enet}, while PatchGAN \cite{GAN} is applied for \textbf{Dnet}.
Within the same epoch, each subnet shares the same parameters, even though it may appear at multiple positions in the training pipeline.

The core point of our scheme is to generate identical guiding images $I_g$s ($I_g$ and $I_g'$) from either cover image $I_o$ or watermarked image $I_w$, ensuring the identical residual can be calculated. To achieve this, we introduce a component \textbf{Guider} to implement guiding strategy, enabling the generation of identical $I_g$s and reversible recovery.

\subsection{Guiding Strategy}
Instead of directly embedding watermark into the cover image, we utilize a guiding image to guide watermark embedding, and the guiding image generated after extraction guides reversible recovery of the cover image. A component \textbf{Guider} is introduced to operate the guiding strategy, and it consists of a \textbf{Gnet} and a \textbf{Hnet} as shown in \cref{flowchart}.

More explicitly, in the embedding phase, cover image $I_o$ is input into \textbf{Guider} to generate guiding image $I_g$, then watermark $wm$ is embedded into $I_g$ with \textbf{Hnet}, and the guiding residual is calculated and transferred to $I_g$. As shown in \cref{guiding}, where $I_g = \textbf{Guider}(I_o)$.
\begin{equation}
	\label{guiding}
	I_w = I_o + (\textbf{H}(I_g||wm) - I_g)
\end{equation}

In the recovery phase after extraction, \textbf{Guider} guarantees that the identical guiding image $I_g'$ can be generated from $I_w$, satisfying $\textbf{Guider}(I_o) = \textbf{Guider}(I_w)$. Then the identical guiding residual can be acquired following the same workflow in the embedding phase, and we can reconstruct $I_o'$ by removing the residual from $I_w$, as shown in \cref{guiding2}, where $I_g' = \textbf{Guider}(I_w)$ and $wm'$ is the extracted watermark.
\begin{equation}
	\label{guiding2}
	I_o' = I_w - (\textbf{H}(I_g'||wm') - I_g')
\end{equation}

The designing of \textbf{Guider} involves two aspects: similarity comparison between $I_g$ and $I_g'$, and consistency comparison of guiding images and original images. The detailed training procedure of \textbf{Guider} is stated in the following part.

\subsection{Training Procedure}
\subsubsection{Stage I} As shown in \cref{flowchart}, Stage I contains two steps: (a) pretraining conventional watermarking subnets, and (b) independently pretraining \textbf{Gnet}. 

(a) In the first step, $I_o$ and $wm$ are concatenated as the input of \textbf{Hnet}, and $I_w$ is the output of \textbf{Hnet}. Then $I_w$ is input into \textbf{Enet} to extract $wm'$. For adversarial training, $I_o$ and $I_w$ are input to \textbf{Dnet} separately to discriminate the local and global imperceptibility of watermark embedding. The loss functions in this step consist of two parts: the hiding loss $\mathcal{L}_H$ and the extraction loss $\mathcal{L}_E$.

The objective of \textbf{Hnet} is to hide $wm$ into $I_o$ unconsciously: $I_w = \textbf{H}(I_o||wm)$, so the visual consistency of $I_o$ and $I_w$ should be guaranteed. Three types of visual consistency loss are applied, i.e., $L2$ loss $\ell_{L2}$, perceptual loss $\ell_{vgg}$, and adversarial loss $\ell_{adv}$.

The $L2$ loss $\ell_{L2}$ measures pixel-level differences of $I_o$ and $I_w$ by calculating $L2$ norm. The perceptual loss $\ell_{vgg}$ measures differences between VGG feature \cite{vgg} of $I_o$ and $I_w$. 
The adversarial loss $\ell_{adv}$ improves the imperceptibility of $I_w$ by leveraging the discriminator \textbf{Dnet} to distinguish cover images and watermarked images, as shown in \cref{adv}.
\begin{equation}
	\label{adv}
	\ell_{adv} = log(\textbf{D}(I_o)) + log(1-\textbf{D}(I_w))
\end{equation}

The objective of \textbf{Enet} is to extract watermark $wm'$ from watermarked image $I_w$, and extract nothing from watermark-free image $I_o$ for copyright validation: $wm'=\textbf{E}(I_w)$. So two types of extraction loss are considered.

The watermark extraction loss $\ell_{wm}$ measures the similarity between the embedded watermark $wm$ and the extracted watermark $wm'$, and we also employ $L2$ norm in this part.
While non-watermark extraction loss $\ell_{nw}$ ensures the network outputs blank images rather than erroneously extracting watermarks from non-watermarked images, another $L2$ norm is calculated between $\textbf{E}(I_o)$ and a blank image $I_{blank}$. In Stage I, the above three subnets are trained collaboratively as the conventional embedding and extraction process, with the loss function $\mathcal{L}_{wm}$ as shown in \cref{L_HE}, where $\eta$ controls the weight of adversarial loss, and $\lambda_1$, $\lambda_2$ controls the relative weights of $\textbf{Hnet}$ and $\textbf{Enet}$.
\begin{equation}
	\label{L_HE}
	\mathcal{L}_{wm} = \lambda_1 (\ell_{L2}+\ell_{vgg})+\lambda_2(\ell_{wm}+ \ell_{nw})+\eta \ell_{adv}
\end{equation}

(b) In the second step of Stage I, \textbf{Gnet} is pretrained to extract maximally identical features from $I_o$ and $I_w$ as the guiding image for reversible embedding. Both the similarity between $I_g$s and the consistency of the output image to the input image are essential, and the former ensures reversibility while the latter improves guidance accuracy. So the loss function of \textbf{Gnet} training contains three parts: the identity guiding loss $\ell_g$ and the consistency guiding loss $\ell_{c}$.

\begin{equation}
	\label{L_G}
	\mathcal{L}_G = \ell_g + \ell_{c}
\end{equation}
The identity guiding loss $\ell_g$ evaluates the pixel-wise identicality of guiding images generated from $I_o$ and $I_w$ with $L2$ norm. The consistency guiding loss $\ell_{c}$ measures the consistency between the guiding images and $I_o$ or $I_w$ to prevent feature degradation, and it consists of two terms, namely the L2 norms of ($I_o, I_g$) and ($I_w, I_g'$).

\subsubsection{Stage II}
All subnets, except for the frozen \textbf{Dnet}, are jointly trained to achieve invisible watermark embedding, efficient extraction, and reversible recovery. As shown in \cref{flowchart}, during the training of \texttt{Hider}, $I_o$ is input into \textbf{Guider}, where it is first processed by \textbf{Gnet} and then embedded a blank watermark by \textbf{Hnet} to generate $I_g$, as shown in \cref{guide}. 
\begin{equation}
	\label{guide}
	I_g = \textbf{H}(\textbf{G}(I_o)||I_{blank})
\end{equation}
Then $I_g$ is concatenated with the watermark $wm$ to generate a watermarked intermediate image $I_t = \textbf{H}(I_g||wm)$. The residual of $I_g$ and $I_t$ is calculated and added to $I_o$ to finish watermark embedding, as shown in \cref{guiding}.

$I_w$ is input into \texttt{Extractor} to extract watermark $wm'$. And then proceeding to reversible recovery training. The guiding image $I_g'$ is generated with $I_w$, following the same process as \cref{guide}, differing only in replacing $I_o$ with $I_w$. Then we compute $I_t'$ and the guiding residual, which is identical to the embedding phase, and the recovered cover image $I_o'$ is calculated as in \cref{guiding2}.

The loss function in Stage II is a combination of loss functions in Stage I, as shown in \cref{S2loss}.

\begin{equation}
	\label{S2loss}
	\mathcal{L}_{C} = \mathcal{L}_{wm} + \lambda_3  \mathcal{L}_G
\end{equation}

\subsection{Robust Extension}
The workflow above serves as a novel deep-model-based reversible watermarking framework, and it neglects to account for watermark robustness. Experimental results show that the current implementation is only robust to slight noise attacks, with detailed experimental data presented in the next section. 

For robustness improving in RRW scenarios, we introduce an additional noise layer \textbf{Noise} after \texttt{Hider} during the training of Stage II to degrade watermarked images before extraction, as shown in \cref{noise}.
\begin{equation}
	\label{noise}
	wm_{noise}' = \textbf{E}(\textbf{Noise}(I_w))
\end{equation}

The noise layer consists of two parallel branches: a JPEG operation and a Gaussian filtering branch. The JPEG branch improves robustness against common distortions such as salt-and-pepper noise, crop, and JPEG compression by implicitly guiding the model to ignore high-frequency components where noise typically occurs. Besides, as noted in \cite{filter}, nonlinear median filtering can be approximated as a linear combination of Gaussian filters. Consequently, Gaussian filtering branch simultaneously trains the model to resist median filtering. Furthermore, since image scaling operations exhibit spectral properties similar to Gaussian filtering and JPEG, it also extends robustness to scaling.

And another loss function $\ell_{noise}$ is introduced to the second term of $\mathcal{L}_{wm}$, as shown in \cref{lossnoise2}, where $\ell_{noise}$ measures the $L2$ norm of $wm$ and $wm_{noise}'$, and $\lambda_2$ is the same weight as \cref{L_HE}. This approach enhances robustness while preserving perfect reversibility.

\begin{equation}
	\label{lossnoise2}
	\mathcal{L}_{wm}' = \mathcal{L}_{wm}+\lambda_2 \ell_{noise}
\end{equation}

\subsection{Reversible Recovery Analysis}
The core points of SiGRRW framework are \emph{identical guiding images generation} and \emph{residual-guided watermarking embedding}. We provide a comprehensive analysis of these points in this part.

\emph{Identical guiding images generation} is fundamental for reversible recovery, and the cover image $I_o$ can be exactly recovered if and only if $I_g=I_g'$ is strictly satisfied. 
Instead of directly using a \textbf{Gnet}, we apply \textbf{Guider}, where an additional \textbf{Hnet} is deployed after \textbf{Gnet}, to ensure identical guiding images generation. 
Because given multiple optimization objectives and loss functions, the model tends to output highly similar yet non-identical results, in other words, the loss function $\ell_g$ yields an extremely small non-zero value, while still introducing non-negligible pixel errors between $I_g$ and $I_g'$: $\textbf{G}(I_w) = \textbf{G}(I_o) + \epsilon$, where $\epsilon$ denotes a minor non-zero error term. 
While removing consistency loss of \textbf{Gnet} will lead to feature degradation of $I_g$ and thereby diminish guidance ability, which will be illustrated in the ablation experiments. So single \textbf{Gnet} is not suitable for guiding image generation.

Given similar UNet network structures, the minor error $\epsilon$ between $I_g$ and $I_g'$ neglected by \textbf{Gnet} will also fall below the response threshold of \textbf{Hnet}, leading both subnets to disregard these perturbations consistently.
An auxiliary \textbf{Hnet} is deployed to compensate for the errors during $I_g$ and $I_g'$ generation, as shown in \cref{Herr}. In other words, the guiding image generation is a two-step process, primary image generation by \textbf{Gnet} and the final guiding image generation by \textbf{Hnet}. More precisely, during training, we set the quantization tolerance $\epsilon$ such that the MSE$(I_g, I_g')\leq 10^{-6}$. Under this condition, the residual error can be completely eliminated by the rounding operation, making $I_gs$ generated by \textbf{Guider} identical. 
\begin{equation}
	\label{Herr}
	\textbf{H}(\textbf{G}(I_o)||I_{blank}) = \textbf{H}((\textbf{G}(I_o) + \epsilon)||I_{blank})
\end{equation}

\par
\emph{Residual-guided watermarking embedding} is another point of SiGRRW, the guiding residual of $I_g$ should be also applicable for embedding $wm$ into $I_o$.
Notably, under the guidance of the consistency guiding loss $\ell_c$, \textbf{Gnet} preserves the majority feature of the input image $I$, and the $L2$ loss $\ell_{L2}$ also constrains \textbf{Hnet} output to retain most structural information. 
As a result, the features representations used for watermark embedding are nearly identical between $I_o$ and $I_g$, allowing the residual to serve as an effective initialization for embedding training.
The collaborative training in Stage II refines embedding residual to better align with the embedding characteristics of $I_o$, thereby ensuring reliable watermark extraction.

\section{Experiments}

\subsection{Basic Setup}
\subsubsection{Dataset}
We use 4000 images from PASCAL VOC \cite{VOC} and LAION-Aesthetics \cite{LAION} for training, which involves both natural images and generated images, and another 200 unseen images for validation. Specifically, \textbf{Hnet}, \textbf{Enet} and \textbf{Dnet} are trained on these images. The output images by \textbf{Hnet} paired with the input ones constitute the training set for \textbf{Gnet}. All the images including watermarks are resized to $256\times 256$, and we apply binary watermark images by default. 
\subsubsection{Implementation Details}
We pretrain subnets \textbf{Hnet}, \textbf{Enet}, \textbf{Dnet} and \textbf{Gnet} for 50 epochs in Stage I, and collaboratively train the entire network for 50 epochs in Stage II, with a batchsize of 4. We apply Adam optimizer with the initial learning-rate = 0.002, and the learning rate decays by 0.2 if the loss does not decrease within 5 epochs. The hyperparameters are set to $(\lambda_1,\lambda_2,\lambda_3) = (10,10,1)$, and $\eta = 2$. We apply a JPEG branch (QF = 50) and a Gaussian Blur branch ($\sigma$ = 7) as the Noise layer.
All experiments are performed using Pytorch 2.7.1 and a NVIDIA 4090 GPU.
\subsubsection{Metrics}
The robustness and imperceptibility of the watermark is assessed, and we evaluate the reversibility by assessing the quality of the recovered images. 
PSNR and SSIM are utilized to evaluate the visual quality and imperceptibility of the watermarked images, and to validate the quality of the recovered image. 
For binary watermarks, we evaluate the accuracy Bit Error Ratio (BER) and ACC = 1 - BER between $wm$ and $wm'$, where BER represents the percentage of incorrect bits.
\subsubsection{Baseline}
We select baselines comprehensively, including both traditional methods and deep learning methods, natural images and model-generated images, and robust watermarking and reversible watermarking methods. And these state-of-the-art (SOTA) watermarking methods include: 
CIN \cite{Ma2022CIN}, MuST \cite{Wang2024Must}, DMIPP \cite{Zhang2022Deep}, IWRN \cite{Kou2025Robust}, RRW-PZMs \cite{Tang2023A}, RRW-FoZM \cite{Fu2023Robust}, DRRW \cite{Chen2025deep} and RRWID\cite{Guo2025RRW}.

\begin{table}[]		
	\centering	
	\small 
	\setlength{\tabcolsep}{1mm}
	\caption{Comparison of basic performance ($256^2$ indicates an image of size $256 \times 256$)}
    \begin{tabular}{@{}l|ccc|c@{}}
        \toprule
        \textbf{Methods} & \textbf{PSNR} & \textbf{SSIM} & \textbf{Payloads} & \textbf{Reversible} \\ 
        \midrule
        CIN \cite{Ma2022CIN}       & 42.56 & 0.9792 & 30 bits& \multirow{4}{*}{\ding{55}} \\
        MuST \cite{Wang2024Must} & 42.89 & 0.9883 & 30 bits          & \\
        DMIPP \cite{Zhang2022Deep} & \textbf{47.29} & \textbf{0.9997} & \textbf{$256^2$ bits} & \\
        IWRN \cite{Kou2025Robust}& 41.55 & 0.9616 & 32 bits          & \\
        \midrule
        RRW-FoZM  \cite{Fu2023Robust}  & 39.43 & 0.9823 & 128 bits         & \multirow{6}{*}{\ding{51}} \\
        RRW-PZMs \cite{Tang2023A}  & 39.34 & 0.9794 & 256 bits         & \\
        DRRW \cite{Chen2025deep} & 40.94 & \textbf{0.9995} & 256 bits         & \\
        RRWID \cite{Guo2025RRW} & 42.48 &0.9850  & 64 bits         & \\
        \textbf{Ours} & \textbf{44.25} & 0.9923 &\textbf{ $256^2$ bits} & \\
        \bottomrule
    \end{tabular}

	\label{comparative}
\end{table}

\begin{table*}[]
	\centering
	\small
	\setlength{\tabcolsep}{1mm}
	\caption{Comparison of robustness results on ACC of watermark extraction under common distortions}
	\begin{tabular}{@{}l|cccccccc@{}}
		\toprule
		\textbf{Method} & \textbf{Gaussian Noise} & \textbf{Gaussian Blur} & \textbf{SaltPepper} & \textbf{JPEG} & \textbf{Median Filter} & \textbf{Scale} & \textbf{Crop} & \textbf{Dropout} \\
		\midrule
		RRW-FoZM & 96.05 & 99.99 & 56.03 & 99.99 & 94.10 & 93.11 & 53.48 & 61.95 \\
		RRW-PZMs & 89.47 & 95.70 & 46.48 & 99.99 & 89.84 & 99.61 & 85.55 & 58.98 \\
		DRRW     & 99.99 & 99.99 & 99.99 & 99.99 & 99.99 & 99.99 & \textbf{99.99} & \textbf{99.99} \\
		\textbf{Ours(standard)}     & 99.87 & 99.84 & 99.82 & 99.11 & 97.45 & 99.81 & 95.20 & 99.69 \\
		\textbf{Ours(256 bit)} & \textbf{100.00} & \textbf{100.00} & \textbf{100.00} & \textbf{100.00} & \textbf{100.00} & \textbf{100.00} & 99.91 & 99.90 \\
		\bottomrule
	\end{tabular}
	\label{comparative_robust}
\end{table*}

\subsection{Comparative Experiment}
To provide an intuitive demonstration of the performance of the proposed scheme, comparative experiments are conducted against several SOTA schemes. Imperceptibility, robustness and embedding capacity are assessed in this part. 256×256 color cover images and 256×256 binary watermarks are employed for the proposed standard scheme.
Considering various embedding patterns of baselines, different watermarking payloads of baselines are applied for comparison. Our framework still maintains outstanding performance even with much higher payloads. 

\subsubsection{Basic Comparison}

\cref{comparative} presents a comparative analysis of visual quality and embedding capacity across SOTA schemes under no-attack conditions. SiGRRW demonstrates outstanding PSNR and SSIM, and achieves the highest PSNR performance among RRW schemes despite much higher payloads than conventional ones, which evidences exceptional imperceptibility of our scheme. 
Furthermore, SiGRRW achieves the highest embedding capacity among all evaluated RRW schemes, establishing a new benchmark for high-capacity RRW.

Typically, deep-model-based watermarking schemes demonstrate superior embedding capacity and imperceptibility but lack reversibility, while conventional RRW schemes show weaker imperceptibility and the reversibility is realized without deep model. Our scheme achieves enhanced embedding capacity and imperceptibility performance while maintaining reversibility using deep model.

\subsubsection{Robustness Comparison}
We evaluate the robustness of our scheme against both common distortions and regeneration attacks. For common distortions, we compare SiGRRW with a series of SOTA RRW schemes that achieve strong robustness under conventional noise. For the more advanced regeneration attacks, prior work has shown that very few watermarking schemes can withstand such attacks, and the only existing RRW method resistant to regeneration is RRWID \cite{Guo2025RRW}, although it is not SOTA against common distortions. So we conduct a dedicated comparison with RRWID under regeneration attacks.

\cref{comparative_robust} presents a comprehensive robustness comparison of RRW schemes under various common distortions. We configure parameters of the attack as follows: Gaussian Noise ($\sigma=0.2$), Salt-and-Pepper Noise (density=0.1), Gaussian Blur (kernel = $7\times7$), Median Filtering (kernel = $7\times7$), JPEG ($QF=50$), Scale (factor=0.5), Crop ($50\times50$), Dropout ($30\%$).
While the robustness of the baseline two-stage schemes relies only upon the robust part, SiGRRW achieves both lossless reversibility and satisfactory robustness with a single watermark with the highest embedding payloads. For the performance of our standard method under additional distortion parameter settings, please refer to \cref{tradeoff}.

Although DRRW shows slightly better robustness than our standard experimental setup, SiGRRW supports a substantially higher embedding capacity. When embedding 256-bit watermark, as shown in \cref{comparative_robust}, our approach still attains SOTA performance. 

We further evaluate the robustness of SiGRRW against VAE-based regeneration attacks. Regeneration attacks \cite{Zhao2024regeneration} built upon VAEs are highly effective at removing most watermarks designed by traditional or deep learning approaches. Since RRWID \cite{Guo2025RRW} is the only existing RRW scheme specifically designed to resist such attacks, we perform a direct comparison with RRWID using the same 64-bit embedding capacity and identical settings. We use the two VAEs with different parameter settings provided by Zhao \cite{Zhao2024regeneration} and apply 60 noise step. The experimental results are presented in \cref{Comprison_regen}, which demonstrates that SiGRRW achieves superior robustness against regeneration attacks compared to RRWID.

\begin{table}[ht] 
	\centering
	\small 
	\setlength{\tabcolsep}{1mm} 
	\caption{BER results under regeneration attacks}
	\begin{tabular}{@{}l|cc|cc@{}}
		\toprule
		VAE Type& \multicolumn{2}{c|}{\textbf{Regen(VAE-Bmshj)}} 
		& \multicolumn{2}{c}{\textbf{Regen(VAE-Cheng)}} \\ 
		\midrule
		Quality& 4 & 5 & 4 & 5 \\ 
		\midrule
		RRWID  & 4.90 & 2.94 & 9.77 & 6.17 \\
		\textbf{Ours(64bit)} & \textbf{4.55} & \textbf{2.10} & \textbf{6.04} & \textbf{1.85} \\
		\bottomrule
	\end{tabular}
	\label{Comprison_regen}
\end{table}

\subsection{Diverse Experiments}
\subsubsection{Recovery after Distortion}
Even when the watermarked image undergoes specific distortion, our scheme actively attempts to recover it to the original one. The recovered image exhibits higher fidelity to the cover image under the same distortion, effectively reducing the perceptual impact of the watermark itself in distorted conditions. \cref{Recovery after Distortion} measures the PSNR between the recovered distorted images and the original images under the same distortion. Our scheme maintains partial reversibility under Gaussian Noise ($\sigma = 0.2$), Gaussian Blur (kernel = $5\times5$), Salt-and-Pepper Noise (density = 0.005) or Median Filtering (kernel = $7\times7$), proactively minimizing watermark induced artifacts in the distorted cover images.

While the other RRW schemes fail to achieve recovery after noise attacks, for the reversible watermark is damaged by distortion and fails to extract auxiliary information after attacks, and the robust one is completely unreversible without auxiliary information. Moreover, the damaged reversible watermark may adversely affect the robust watermark and the cover image itself, leading to further degradation in fidelity.

\begin{table}[ht]
	\centering
    \small 
    \setlength{\tabcolsep}{1mm} 
    \caption{PSNR results of recovered images after distortion}
    \begin{tabular}{@{}l|cccc@{}}
        \toprule
        & \textbf{GausNoise} & \textbf{GausBlur} & \textbf{SaltPepper} & \textbf{Median Filter} \\ 
        \midrule
        $(I_w,I_o)$ & 40.80 & 42.52 & 20.73 & 42.76 \\
        $(I'_o,I_o)$ & 50.63 & 47.20 & 21.78 & 44.86 \\
        \bottomrule
    \end{tabular}	
	\label{Recovery after Distortion}
\end{table}

\subsubsection{Tradeoff between Robustness and Reversibility}
\begin{figure}[t]
	\centering	
	\begin{minipage}{0.23\textwidth}
		\centering
		\includegraphics[width=\linewidth]{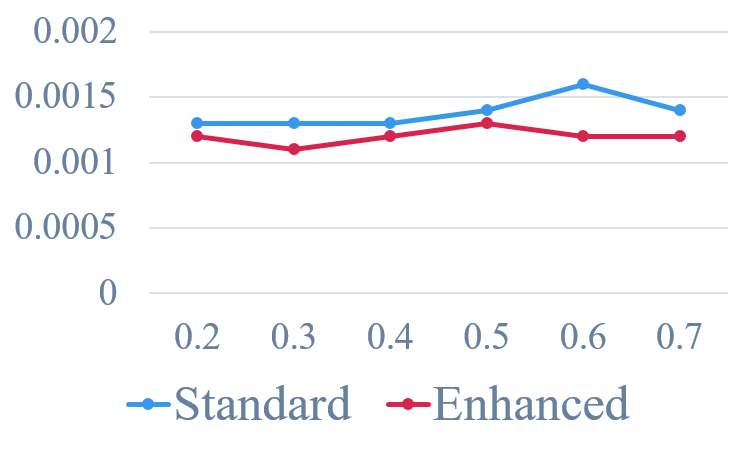}
		\subcaption{Gaussian Noise}
		\label{tfig:sub1}
	\end{minipage}
	\begin{minipage}{0.23\textwidth}
		\centering
		\includegraphics[width=\linewidth]{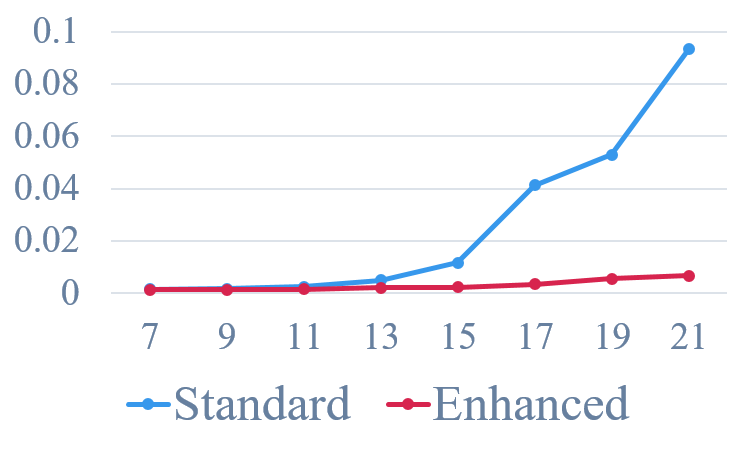}
		\subcaption{Gaussian Blur}
		\label{tfig:sub2}
	\end{minipage}	
	\begin{minipage}{0.23\textwidth}
		\centering
		\includegraphics[width=\linewidth]{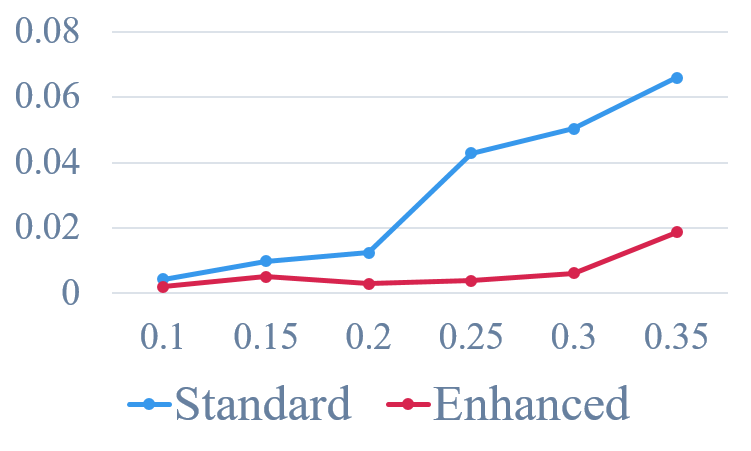}
		\subcaption{Salt-and-Pepper Noise}
		\label{tfig:sub3}
	\end{minipage}
	\begin{minipage}{0.23\textwidth}
		\centering
		\includegraphics[width=\linewidth]{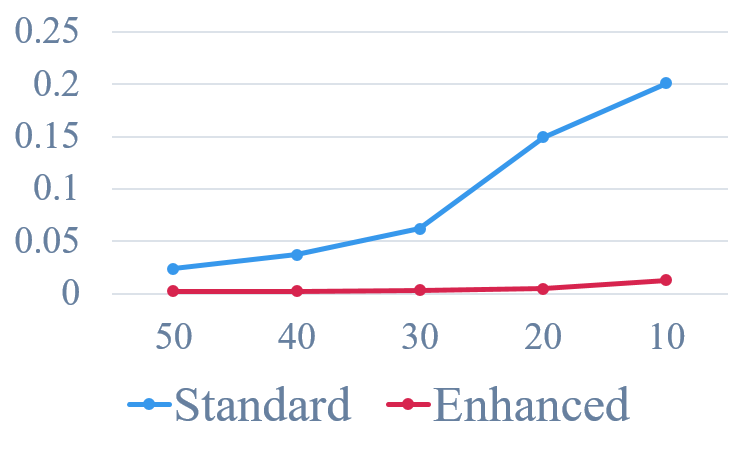}
		\subcaption{JPEG}
		\label{tfig:sub4}
	\end{minipage}
	\begin{minipage}{0.23\textwidth}
		\centering
		\includegraphics[width=\linewidth]{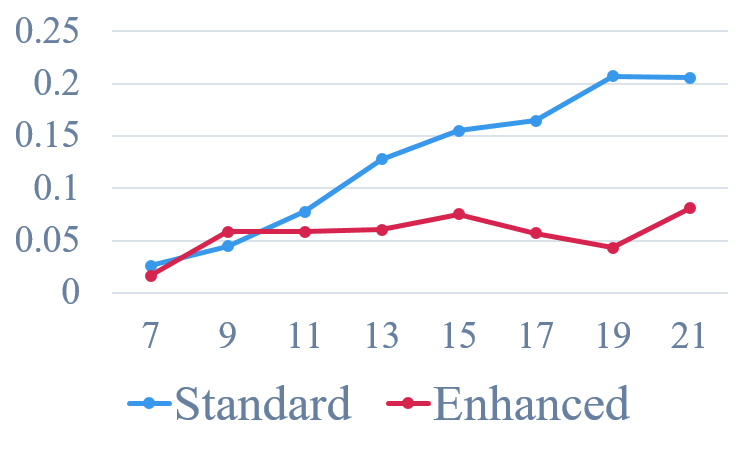}
		\subcaption{Median Filtering}
		\label{tfig:sub5}
	\end{minipage}
	\begin{minipage}{0.23\textwidth}
		\centering
		\includegraphics[width=\linewidth]{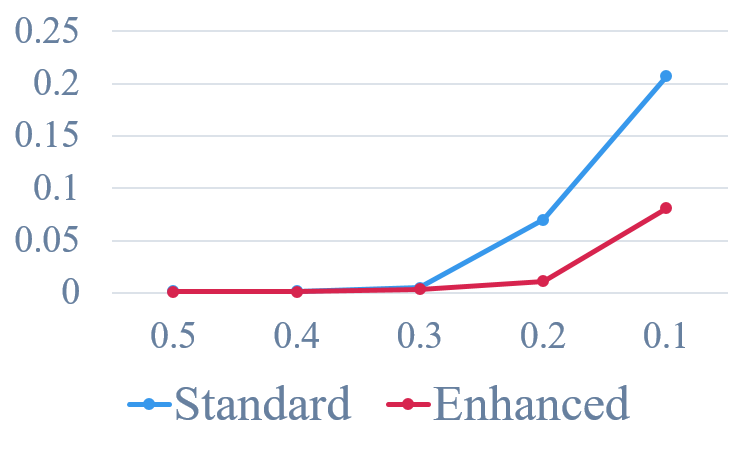}
		\subcaption{Scaling}
		\label{tfig:sub6}
	\end{minipage}

	\caption{Robust Enhancing Results on BER}
	\label{tradeoff}
\end{figure}

In certain scenarios involving severe distortion, our framework can enhance robustness with partial sacrifice of reversibility. Since the image is damaged by external distortions, lossless recovery has failed, and the proposed watermark essentially serves as a robust watermark. So we dynamically prioritize the strength of noise layer to boost watermark robustness while preserving partial reversibility. For the robustness enhanced version, we set the weight of noise layer as 5, and the strength of the simulation distortion as JPEG ($QF=10$) and Gaussian Blur ($\sigma = 15$).
As shown in \cref{tradeoff}, we can reduce the BER of the enhanced scheme in severe distortion scenarios compared with the standard one, achieving PSNR($I_o, I_w$) = 38.53 and PSNR($I_o, I_o'$) = 56.18.

\begin{table}[]
	\centering
	\small
	\setlength{\tabcolsep}{1mm}
	\caption{PSNR of ablation experiments of Guider}
	\begin{tabular}{@{}l|cccc@{}}
		\toprule
		& \textbf{with all} & \textbf{w/o Hnet} & \textbf{w/o Gnet} & \textbf{w/o $\ell_c$} \\ 
		\midrule
		$(I_g, I_g')$ & $\infty$ & 64.73 & 77.54 & $\infty$ \\
		$(I_o, I_w)$  & 44.25 & 45.91 & 34.43 & 28.56 \\
		$(I_o, I_o')$ & $\infty$ & 77.96 & 78.58 & 66.21 \\
		\bottomrule
	\end{tabular}
	\label{Ablation1}
\end{table}

\subsection{Ablation Study}

\subsubsection{Ablation of \textbf{Guide}}
We conduct ablation study on the components of \textbf{Guide}. Several PSNR values are presented in \cref{Ablation1} to measure the influence of relevant components on performance.
The absence of either \textbf{Gnet} or \textbf{Hnet} in \textbf{Guide} leads to non-negligible errors in guiding images generation as shown in Row 1, and then errors propagate into the final recovered images as shown in Row 3. So both components require mutual compensation to address their respective errors in guiding image generation.

The consistency guiding loss $\ell_{c}$ is crucial for preserving the semantic information of the cover image. As illustrated in \cref{Ablation1}, without $\ell_{c}$, the generation of $I_w$ exhibits feature degradation, which severely compromises imperceptibility.

\subsubsection{Ablation of the Noise Layer}
As shown in \cref{Ablation2}, removing either of the noise attack simulation leads to weaker robustness to certain common distortions. Training without \textbf{Noise} weakens the robustness, it can resist only limited distortion while failing against most attacks, resulting in verification failures. And removal of JPEG directly compromises robustness against JPEG compression, while removing Filtering leads to significant degradation in robustness against Gaussian blur and median filtering.

\begin{table}[t]	
	\centering
		\small 
		\setlength{\tabcolsep}{1mm}
		\caption{Extraction ACC with/without Noise Layer. "0" means a blank watermark is extracted}
		\begin{tabular}{@{}l|cccc@{}}
			\toprule
			\textbf{Noise Layer} & \textbf{with all} & \textbf{w/o JPEG} & \textbf{w/o Filtering} & \textbf{w/o \textbf{Noise}} \\
			\midrule
			Gaussian Noise  & 99.87 & 99.77 & 99.89 & 99.78 \\
			Gaussian Blur   & 99.84 & 99.50 & 22.87 & 0     \\
			SaltPepper & 99.82 & 99.54 & 99.29 & 98.32 \\
			JPEG       & 99.11 & 0     & 74.03 & 0     \\
			Median Filter  & 97.45 & 90.19 & 17.52 & 0     \\
			Scale      & 99.81 & 98.74 & 99.83 & 0     \\
			Crop    & 95.20 & 86.90 & 99.43 & 74.60 \\
			Dropout    & 99.69 & 99.42 & 99.26 & 95.59 \\
			\bottomrule
		\end{tabular}
	
	\label{Ablation2}
\end{table}

\subsubsection{Ablation of Subnets}
We conduct ablation study on several components of the subnets including \textbf{Hnet}, \textbf{Dnet} and \textbf{Gnet}.

The loss functions of \textbf{Hnet} during embedding training process are analyzed. Both $\ell_{L2}$ and $\ell_{vgg}$ contribute positively to the watermark imperceptibility, as illustrated in the first part of \cref{Ablation_HD}. Removing either loss function results in degradation of PSNR. A separate ablation study will be conducted on loss function $\ell_{adv}$ in the following.

The weight of \textbf{Dnet} is analyzed. As shown in the second part of \cref{Ablation_HD}, where $\eta$ is the weight of \textbf{Dnet}. It serves as the discriminator in our framework, specializing in detecting watermark presence within images. Ablating \textbf{Dnet} during training leads to degradation in watermark imperceptibility and reversibility.  Extremely high weight of \textbf{Dnet} results in excessive bias to undetectability, and induces low PSNR of the watermarked image, high BER of extracted watermark and also compromises reversibility.

\begin{table}[t!]
	\centering
	\small 
	\setlength{\tabcolsep}{1mm}
	\caption{Ablation study results of subnets}
	\begin{tabular}{@{}l|l|cccc@{}}
		\toprule
		Subnet                & parameter                       & PSNR($I_o,I_w$)  & SSIM   & PSNR($I_o,I_o'$) & BER   \\ \midrule
		\multirow{3}{*}{\textbf{Hnet}} & w/o $\ell_{L2}$  & 40.73 & 0.9885 & $\infty$ & 0.000 \\
		& w/o $\ell_{vgg}$ & 39.20 & 0.9799 & $\infty$   & 0.000 \\
		& with all   & \textbf{44.56} & \textbf{0.9903} & $\infty$   & 0.000 \\ 
		\midrule
		\multirow{6}{*}{\textbf{Dnet}} & $\eta=0$       & 28.33 & 0.8967 & 45.98    & 0.013 \\
		& $\eta=1$        & 36.03 & 0.7940 & $\infty$ & 0.005 \\
		& $\eta=2$      & \textbf{44.56} & \textbf{0.9903} & $\infty$ & \textbf{0.000} \\
		& $\eta=3$       & 42.17 & 0.9801 & $\infty$ & 0.000 \\
		& $\eta=5$       & 44.50 & 0.9874 & 96.67    & 0.002 \\
		& $\eta=10$       & 27.37 & 0.8226 & 50.48    & 0.790 \\ 
		\midrule
		\multirow{4}{*}{\textbf{Gnet}} & layer = 4       & 31.62 & 0.8917 & 78.10   &0.001 \\
		& layer = 5        & \textbf{44.56} & \textbf{0.9903} & \textbf{$\infty$} & \textbf{0.000} \\
		& layer = 6      & 30.98 & 0.8941 & $\infty$ & 0.001 \\
		\bottomrule
	\end{tabular}

	\label{Ablation_HD}
\end{table}

The layers of \textbf{Gnet} are studied. Given 7 layers of \textbf{Hnet} for embedding, \textbf{Gnet} with 5 layers is sufficient to satisfy identical guiding images generation requirements as shown in the third part of \cref{Ablation_HD}. Increasing the layer count unnecessarily prolongs training costs, while insufficient layers result in failure of identical guiding images generation and reduced watermark verification success rates.

\section{Conclusion}
In this paper, we propose SiGRRW, a novel single-watermark Robust Reversible Watermarking (RRW) framework. 
Apart from conventional two-stage RRW schemes, we achieve simultaneous robustness and reversibility with only one watermark, significantly enhancing watermark imperceptibility and embedding capacity. \textbf{Guider} for guiding image generation is introduced to guide reversible watermark embedding via embedding residual. SiGRRW achieves satisfactory results against a wide range of common distortions as well as regeneration attacks. Extensive experimental results demonstrate that our scheme exhibits outstanding imperceptibility and robustness performance with large capacity, while maintaining reversibility.

{
    \small
    \bibliographystyle{ieeenat_fullname}
    \bibliography{main}

@article{Ni2006,
	author       = {Zhicheng Ni and
	Yun{-}Qing Shi and
	Nirwan Ansari and
	Wei Su},
	title        = {Reversible data hiding},
	journal      = {{IEEE} Trans. Circuits Syst. Video Technol.},
	volume       = {16},
	number       = {3},
	pages        = {354--362},
	year         = {2006},
	url          = {https://doi.org/10.1109/TCSVT.2006.869964},
	doi          = {10.1109/TCSVT.2006.869964},
	bibsource    = {dblp computer science bibliography, https://dblp.org}
}

@article{Tian2003,
	author       = {Jun Tian},
	title        = {Reversible data embedding using a difference expansion},
	journal      = {{IEEE} Trans. Circuits Syst. Video Technol.},
	volume       = {13},
	number       = {8},
	pages        = {890--896},
	year         = {2003},
	url          = {https://doi.org/10.1109/TCSVT.2003.815962},
	doi          = {10.1109/TCSVT.2003.815962},
	bibsource    = {dblp computer science bibliography, https://dblp.org}
}

@article{Sachnev2009,
	author       = {Vasiliy Sachnev and
	Hyoung{-}Joong Kim and
	Jeho Nam and
	Suresh Sundaram and
	Yun{-}Qing Shi},
	title        = {Reversible Watermarking Algorithm Using Sorting and Prediction},
	journal      = {{IEEE} Trans. Circuits Syst. Video Technol.},
	volume       = {19},
	number       = {7},
	pages        = {989--999},
	year         = {2009},
	url          = {https://doi.org/10.1109/TCSVT.2009.2020257},
	doi          = {10.1109/TCSVT.2009.2020257},
	bibsource    = {dblp computer science bibliography, https://dblp.org}
}

@article{Wang2020,
	author       = {Xiang Wang and
	Xiaolong Li and
	Qingqi Pei},
	title        = {Independent Embedding Domain Based Two-Stage Robust Reversible Watermarking},
	journal      = {{IEEE} Trans. Circuits Syst. Video Technol.},
	volume       = {30},
	number       = {8},
	pages        = {2406--2417},
	year         = {2020},
	url          = {https://doi.org/10.1109/TCSVT.2019.2915116},
	doi          = {10.1109/TCSVT.2019.2915116},
	bibsource    = {dblp computer science bibliography, https://dblp.org}
}

@article{Xiong2022Robust,
	author       = {Lizhi Xiong and
	Xiao Han and
	Ching{-}Nung Yang and
	Yun{-}Qing Shi},
	title        = {Robust Reversible Watermarking in Encrypted Image With Secure Multi-Party
	Based on Lightweight Cryptography},
	journal      = {{IEEE} Trans. Circuits Syst. Video Technol.},
	volume       = {32},
	number       = {1},
	pages        = {75--91},
	year         = {2022},
	url          = {https://doi.org/10.1109/TCSVT.2021.3055072},
	doi          = {10.1109/TCSVT.2021.3055072},
	bibsource    = {dblp computer science bibliography, https://dblp.org}
}

@article{Tang2023A,
	author       = {Yichao Tang and
	Shuai Wang and
	Chuntao Wang and
	Shijun Xiang and
	Yiu{-}Ming Cheung},
	title        = {A Highly Robust Reversible Watermarking Scheme Using Embedding Optimization
	and Rounded Error Compensation},
	journal      = {{IEEE} Trans. Circuits Syst. Video Technol.},
	volume       = {33},
	number       = {4},
	pages        = {1593--1609},
	year         = {2023},
	url          = {https://doi.org/10.1109/TCSVT.2022.3216849},
	doi          = {10.1109/TCSVT.2022.3216849},
	bibsource    = {dblp computer science bibliography, https://dblp.org}
}

@article{Wang2024,
	author       = {Haodong Wang and
	Heng Yao and
	Chuan Qin and
	Xinpeng Zhang},
	title        = {When Robust Reversible Watermarking Meets Cropping Attacks},
	journal      = {{IEEE} Trans. Circuits Syst. Video Technol.},
	volume       = {34},
	number       = {12},
	pages        = {13282--13296},
	year         = {2024},
	url          = {https://doi.org/10.1109/TCSVT.2024.3443315},
	doi          = {10.1109/TCSVT.2024.3443315},
	bibsource    = {dblp computer science bibliography, https://dblp.org}
}

@article{Chen2025deep,
	author       = {Jiale Chen and
	Wei Wang and
	Chongyang Shi and
	Li Dong and
	Yuanman Li and
	Xiping Hu},
	title        = {Deep Robust Reversible Watermarking},
	journal      = {CoRR},
	volume       = {abs/2503.02490},
	year         = {2025},
	url          = {https://doi.org/10.48550/arXiv.2503.02490},
	doi          = {10.48550/ARXIV.2503.02490},
	eprinttype    = {arXiv},
	eprint       = {2503.02490},
	bibsource    = {dblp computer science bibliography, https://dblp.org}
}

@article{An2012Robust,
	author       = {Lingling An and
	Xinbo Gao and
	Xuelong Li and
	Dacheng Tao and
	Cheng Deng and
	Jie Li},
	title        = {Robust Reversible Watermarking via Clustering and Enhanced Pixel-Wise
	Masking},
	journal      = {{IEEE} Trans. Image Process.},
	volume       = {21},
	number       = {8},
	pages        = {3598--3611},
	year         = {2012},
	url          = {https://doi.org/10.1109/TIP.2012.2191564},
	doi          = {10.1109/TIP.2012.2191564},
	bibsource    = {dblp computer science bibliography, https://dblp.org}
}

@article{Zhang2022Deep,
	author       = {Jie Zhang and
	Dongdong Chen and
	Jing Liao and
	Weiming Zhang and
	Huamin Feng and
	Gang Hua and
	Nenghai Yu},
	title        = {Deep Model Intellectual Property Protection via Deep Watermarking},
	journal      = {{IEEE} Trans. Pattern Anal. Mach. Intell.},
	volume       = {44},
	number       = {8},
	pages        = {4005--4020},
	year         = {2022},
	url          = {https://doi.org/10.1109/TPAMI.2021.3064850},
	doi          = {10.1109/TPAMI.2021.3064850},
	bibsource    = {dblp computer science bibliography, https://dblp.org}
}

@inproceedings{Fernandez2023Stable,
	author       = {Pierre Fernandez and
	Guillaume Couairon and
	Herv{\'{e}} J{\'{e}}gou and
	Matthijs Douze and
	Teddy Furon},
	title        = {The Stable Signature: Rooting Watermarks in Latent Diffusion Models},
	booktitle    = {{IEEE/CVF} International Conference on Computer Vision, {ICCV} 2023,
	Paris, France, October 1-6, 2023},
	pages        = {22409--22420},
	publisher    = {{IEEE}},
	year         = {2023},
	url          = {https://doi.org/10.1109/ICCV51070.2023.02053},
	doi          = {10.1109/ICCV51070.2023.02053},
	bibsource    = {dblp computer science bibliography, https://dblp.org}
}

@inproceedings{Yu2021Artificial,
	author       = {Ning Yu and
	Vladislav Skripniuk and
	Sahar Abdelnabi and
	Mario Fritz},
	title        = {Artificial Fingerprinting for Generative Models: Rooting Deepfake
	Attribution in Training Data},
	booktitle    = {2021 {IEEE/CVF} International Conference on Computer Vision, {ICCV}
	2021, Montreal, QC, Canada, October 10-17, 2021},
	pages        = {14428--14437},
	publisher    = {{IEEE}},
	year         = {2021},
	url          = {https://doi.org/10.1109/ICCV48922.2021.01418},
	doi          = {10.1109/ICCV48922.2021.01418},
	bibsource    = {dblp computer science bibliography, https://dblp.org}
}

@inproceedings{Yu2022Responsible,
	author       = {Ning Yu and
	Vladislav Skripniuk and
	Dingfan Chen and
	Larry S. Davis and
	Mario Fritz},
	title        = {Responsible Disclosure of Generative Models Using Scalable Fingerprinting},
	booktitle    = {The Tenth International Conference on Learning Representations, {ICLR}
	2022, Virtual Event, April 25-29, 2022},
	publisher    = {OpenReview.net},
	year         = {2022},
	url          = {https://openreview.net/forum?id=sOK-zS6WHB},
	bibsource    = {dblp computer science bibliography, https://dblp.org}
}

@inproceedings{Kou2025Robust,
	author       = {Feifei Kou and
	Yuhan Yao and
	Siyuan Yao and
	Jiahao Wang and
	Lei Shi and
	Yawen Li and
	Xuejing Kang},
	editor       = {Toby Walsh and
	Julie Shah and
	Zico Kolter},
	title        = {{IWRN:} {A} Robust Blind Watermarking Method for Artwork Image Copyright
	Protection Against Noise Attack},
	booktitle    = {AAAI-25, Sponsored by the Association for the Advancement of Artificial
	Intelligence, February 25 - March 4, 2025, Philadelphia, PA, {USA}},
	pages        = {370--378},
	publisher    = {{AAAI} Press},
	year         = {2025},
	url          = {https://doi.org/10.1609/aaai.v39i1.32015},
	doi          = {10.1609/AAAI.V39I1.32015},
	bibsource    = {dblp computer science bibliography, https://dblp.org}
}

@inproceedings{Wang2024MuST,
	author       = {Guanjie Wang and
	Zehua Ma and
	Chang Liu and
	Xi Yang and
	Han Fang and
	Weiming Zhang and
	Nenghai Yu},
	editor       = {Michael J. Wooldridge and
	Jennifer G. Dy and
	Sriraam Natarajan},
	title        = {MuST: Robust Image Watermarking for Multi-Source Tracing},
	booktitle    = {Thirty-Eighth {AAAI} Conference on Artificial Intelligence, {AAAI}
	2024, Thirty-Sixth Conference on Innovative Applications of Artificial
	Intelligence, {IAAI} 2024, Fourteenth Symposium on Educational Advances
	in Artificial Intelligence, {EAAI} 2014, February 20-27, 2024, Vancouver,
	Canada},
	pages        = {5364--5371},
	publisher    = {{AAAI} Press},
	year         = {2024},
	url          = {https://doi.org/10.1609/aaai.v38i6.28344},
	doi          = {10.1609/AAAI.V38I6.28344},
	bibsource    = {dblp computer science bibliography, https://dblp.org}
}

@inproceedings{Fang2025CoSDA,
	author       = {Han Fang and
	Kejiang Chen and
	Zijin Yang and
	Bosen Cui and
	Weiming Zhang and
	Ee{-}Chien Chang},
	editor       = {Toby Walsh and
	Julie Shah and
	Zico Kolter},
	title        = {CoSDA: Enhancing the Robustness of Inversion-based Generative Image
	Watermarking Framework},
	booktitle    = {AAAI-25, Sponsored by the Association for the Advancement of Artificial
	Intelligence, February 25 - March 4, 2025, Philadelphia, PA, {USA}},
	pages        = {2888--2896},
	publisher    = {{AAAI} Press},
	year         = {2025},
	url          = {https://doi.org/10.1609/aaai.v39i3.32295},
	doi          = {10.1609/AAAI.V39I3.32295},
	bibsource    = {dblp computer science bibliography, https://dblp.org}
}

@article{Zhang2024Robust,
	author       = {Jie Zhang and
	Dongdong Chen and
	Jing Liao and
	Zehua Ma and
	Han Fang and
	Weiming Zhang and
	Huamin Feng and
	Gang Hua and
	Nenghai Yu},
	title        = {Robust Model Watermarking for Image Processing Networks via Structure
	Consistency},
	journal      = {{IEEE} Trans. Pattern Anal. Mach. Intell.},
	volume       = {46},
	number       = {10},
	pages        = {6985--6992},
	year         = {2024},
	url          = {https://doi.org/10.1109/TPAMI.2024.3381543},
	doi          = {10.1109/TPAMI.2024.3381543},
	bibsource    = {dblp computer science bibliography, https://dblp.org}
}

@inproceedings{Yang2024Saussian,
	author       = {Zijin Yang and
	Kai Zeng and
	Kejiang Chen and
	Han Fang and
	Weiming Zhang and
	Nenghai Yu},
	title        = {Gaussian Shading: Provable Performance-Lossless Image Watermarking
	for Diffusion Models},
	booktitle    = {{IEEE/CVF} Conference on Computer Vision and Pattern Recognition,
	{CVPR} 2024, Seattle, WA, USA, June 16-22, 2024},
	pages        = {12162--12171},
	publisher    = {{IEEE}},
	year         = {2024},
	url          = {https://doi.org/10.1109/CVPR52733.2024.01156},
	doi          = {10.1109/CVPR52733.2024.01156},
	bibsource    = {dblp computer science bibliography, https://dblp.org}
}

@article{Lin2025CycleGAN,
	author       = {Dongdong Lin and
	Benedetta Tondi and
	Bin Li and
	Mauro Barni},
	title        = {A CycleGAN Watermarking Method for Ownership Verification},
	journal      = {{IEEE} Trans. Dependable Secur. Comput.},
	volume       = {22},
	number       = {2},
	pages        = {1040--1054},
	year         = {2025},
	url          = {https://doi.org/10.1109/TDSC.2024.3424900},
	doi          = {10.1109/TDSC.2024.3424900},
	bibsource    = {dblp computer science bibliography, https://dblp.org}
}

@article{Wu2021watermarking,
	author       = {Hanzhou Wu and
	Gen Liu and
	Yuwei Yao and
	Xinpeng Zhang},
	title        = {Watermarking Neural Networks With Watermarked Images},
	journal      = {{IEEE} Trans. Circuits Syst. Video Technol.},
	volume       = {31},
	number       = {7},
	pages        = {2591--2601},
	year         = {2021},
	url          = {https://doi.org/10.1109/TCSVT.2020.3030671},
	doi          = {10.1109/TCSVT.2020.3030671},
	bibsource    = {dblp computer science bibliography, https://dblp.org}
}

@article{Fu2023Robust,
	author       = {Dahao Fu and
	Xiaoyi Zhou and
	Liaoran Xu and
	Kaiyue Hou and
	Xianyi Chen},
	title        = {Robust Reversible Watermarking by Fractional Order Zernike Moments
	and Pseudo-Zernike Moments},
	journal      = {{IEEE} Trans. Circuits Syst. Video Technol.},
	volume       = {33},
	number       = {12},
	pages        = {7310--7326},
	year         = {2023},
	url          = {https://doi.org/10.1109/TCSVT.2023.3279116},
	doi          = {10.1109/TCSVT.2023.3279116},
	bibsource    = {dblp computer science bibliography, https://dblp.org}
}

@inproceedings{Ma2022CIN,
	author       = {Rui Ma and
	Mengxi Guo and
	Yi Hou and
	Fan Yang and
	Yuan Li and
	Huizhu Jia and
	Xiaodong Xie},
	editor       = {Jo{\~{a}}o Magalh{\~{a}}es and
	Alberto Del Bimbo and
	Shin'ichi Satoh and
	Nicu Sebe and
	Xavier Alameda{-}Pineda and
	Qin Jin and
	Vincent Oria and
	Laura Toni},
	title        = {Towards Blind Watermarking: Combining Invertible and Non-invertible
	Mechanisms},
	booktitle    = {{MM} '22: The 30th {ACM} International Conference on Multimedia, Lisboa,
	Portugal, October 10 - 14, 2022},
	pages        = {1532--1542},
	publisher    = {{ACM}},
	year         = {2022},
	url          = {https://doi.org/10.1145/3503161.3547950},
	doi          = {10.1145/3503161.3547950},
	bibsource    = {dblp computer science bibliography, https://dblp.org}
}

@article{VOC,
	author       = {Mark Everingham and
	Luc Van Gool and
	Christopher K. I. Williams and
	John M. Winn and
	Andrew Zisserman},
	title        = {The Pascal Visual Object Classes {(VOC)} Challenge},
	journal      = {Int. J. Comput. Vis.},
	volume       = {88},
	number       = {2},
	pages        = {303--338},
	year         = {2010},
	url          = {https://doi.org/10.1007/s11263-009-0275-4},
	doi          = {10.1007/S11263-009-0275-4},
	bibsource    = {dblp computer science bibliography, https://dblp.org}
}

@inproceedings{vgg,
	title={Perceptual losses for real-time style transfer and super-resolution},
	author={Johnson, Justin and Alahi, Alexandre and Fei-Fei, Li},
	booktitle={Computer Vision--ECCV 2016: 14th European Conference, Amsterdam, The Netherlands, October 11-14, 2016, Proceedings, Part II 14},
	pages={694--711},
	year={2016},
	organization={Springer}
}

@inproceedings{unet,
	author       = {Olaf Ronneberger and
	Philipp Fischer and
	Thomas Brox},
	editor       = {Nassir Navab and
	Joachim Hornegger and
	William M. Wells III and
	Alejandro F. Frangi},
	title        = {U-Net: Convolutional Networks for Biomedical Image Segmentation},
	booktitle    = {Medical Image Computing and Computer-Assisted Intervention - {MICCAI}
	2015 - 18th International Conference Munich, Germany, October 5 -
	9, 2015, Proceedings, Part {III}},
	series       = {Lecture Notes in Computer Science},
	volume       = {9351},
	pages        = {234--241},
	publisher    = {Springer},
	year         = {2015},
	url          = {https://doi.org/10.1007/978-3-319-24574-4\_28},
	doi          = {10.1007/978-3-319-24574-4\_28},
	bibsource    = {dblp computer science bibliography, https://dblp.org}
}

@inproceedings{CEILnet,
	author       = {Qingnan Fan and
	Jiaolong Yang and
	Gang Hua and
	Baoquan Chen and
	David P. Wipf},
	title        = {A Generic Deep Architecture for Single Image Reflection Removal and
	Image Smoothing},
	booktitle    = {{IEEE} International Conference on Computer Vision, {ICCV} 2017, Venice,
	Italy, October 22-29, 2017},
	pages        = {3258--3267},
	publisher    = {{IEEE} Computer Society},
	year         = {2017},
	url          = {https://doi.org/10.1109/ICCV.2017.351},
	doi          = {10.1109/ICCV.2017.351},
	bibsource    = {dblp computer science bibliography, https://dblp.org}
}

@inproceedings{GAN,
	author       = {Phillip Isola and
	Jun{-}Yan Zhu and
	Tinghui Zhou and
	Alexei A. Efros},
	title        = {Image-to-Image Translation with Conditional Adversarial Networks},
	booktitle    = {2017 {IEEE} Conference on Computer Vision and Pattern Recognition,
	{CVPR} 2017, Honolulu, HI, USA, July 21-26, 2017},
	pages        = {5967--5976},
	publisher    = {{IEEE} Computer Society},
	year         = {2017},
	url          = {https://doi.org/10.1109/CVPR.2017.632},
	doi          = {10.1109/CVPR.2017.632},
	bibsource    = {dblp computer science bibliography, https://dblp.org}
}

@ARTICLE{filter,
	author={Bovik, A. and Huang, T. and Munson, D.},
	journal={IEEE Transactions on Acoustics, Speech, and Signal Processing}, 
	title={A generalization of median filtering using linear combinations of order statistics}, 
	year={1983},
	volume={31},
	number={6},
	pages={1342-1350},
	doi={10.1109/TASSP.1983.1164247}}

@inproceedings{PlugandPlay,
	author       = {Xun Xian and
	Ganghua Wang and
	Xuan Bi and
	Jayanth Srinivasa and
	Ashish Kundu and
	Mingyi Hong and
	Jie Ding},
	editor       = {Amir Globersons and
	Lester Mackey and
	Danielle Belgrave and
	Angela Fan and
	Ulrich Paquet and
	Jakub M. Tomczak and
	Cheng Zhang},
	title        = {{RAW:} {A} Robust and Agile Plug-and-Play Watermark Framework for
	AI-Generated Images with Provable Guarantees},
	booktitle    = {Advances in Neural Information Processing Systems 38: Annual Conference
	on Neural Information Processing Systems 2024, NeurIPS 2024, Vancouver,
	BC, Canada, December 10 - 15, 2024},
	year         = {2024},
	url          = {http://papers.nips.cc/paper\_files/paper/2024/hash/ee62ab636066cf45a27246acca9545b7-Abstract-Conference.html},
	bibsource    = {dblp computer science bibliography, https://dblp.org}
}

@article{Wan2022,
	author       = {Wenbo Wan and
	Jun Wang and
	Yunming Zhang and
	Jing Li and
	Hui Yu and
	Jiande Sun},
	title        = {A comprehensive survey on robust image watermarking},
	journal      = {Neurocomputing},
	volume       = {488},
	pages        = {226--247},
	year         = {2022},
	url          = {https://doi.org/10.1016/j.neucom.2022.02.083},
	doi          = {10.1016/J.NEUCOM.2022.02.083},
	bibsource    = {dblp computer science bibliography, https://dblp.org}
}

@inproceedings{LAION,
	author       = {Christoph Schuhmann and
	Romain Beaumont and
	Richard Vencu and
	Cade Gordon and
	Ross Wightman and
	Mehdi Cherti and
	Theo Coombes and
	Aarush Katta and
	Clayton Mullis and
	Mitchell Wortsman and
	Patrick Schramowski and
	Srivatsa Kundurthy and
	Katherine Crowson and
	Ludwig Schmidt and
	Robert Kaczmarczyk and
	Jenia Jitsev},
	editor       = {Sanmi Koyejo and
	S. Mohamed and
	A. Agarwal and
	Danielle Belgrave and
	K. Cho and
	A. Oh},
	title        = {{LAION-5B:} An open large-scale dataset for training next generation
	image-text models},
	booktitle    = {Advances in Neural Information Processing Systems 35: Annual Conference
	on Neural Information Processing Systems 2022, NeurIPS 2022, New Orleans,
	LA, USA, November 28 - December 9, 2022},
	year         = {2022},
	url          = {http://papers.nips.cc/paper\_files/paper/2022/hash/a1859debfb3b59d094f3504d5ebb6c25-Abstract-Datasets\_and\_Benchmarks.html},
	bibsource    = {dblp computer science bibliography, https://dblp.org}
}

@inproceedings{Zhao2024regeneration,
	author       = {Xuandong Zhao and
	Kexun Zhang and
	Zihao Su and
	Saastha Vasan and
	Ilya Grishchenko and
	Christopher Kruegel and
	Giovanni Vigna and
	Yu{-}Xiang Wang and
	Lei Li},
	editor       = {Amir Globersons and
	Lester Mackey and
	Danielle Belgrave and
	Angela Fan and
	Ulrich Paquet and
	Jakub M. Tomczak and
	Cheng Zhang},
	title        = {Invisible Image Watermarks Are Provably Removable Using Generative
	{AI}},
	booktitle    = {Advances in Neural Information Processing Systems 38: Annual Conference
	on Neural Information Processing Systems 2024, NeurIPS 2024, Vancouver,
	BC, Canada, December 10 - 15, 2024},
	year         = {2024},
	url          = {http://papers.nips.cc/paper\_files/paper/2024/hash/10272bfd0371ef960ec557ed6c866058-Abstract-Conference.html},
	bibsource    = {dblp computer science bibliography, https://dblp.org}
}

@article{Guo2025RRW,
	author       = {Bobiao Guo and
	Ping Ping and
	Fan Liu and
	Feng Xu},
	title        = {Robust Reversible Watermarking With Invisible Distortion Against {VAE}
	Watermark Removal},
	journal      = {{IEEE} Trans. Image Process.},
	volume       = {34},
	pages        = {6386--6401},
	year         = {2025},
	url          = {https://doi.org/10.1109/TIP.2025.3613958},
	doi          = {10.1109/TIP.2025.3613958},
	bibsource    = {dblp computer science bibliography, https://dblp.org}
}

@inproceedings{YangXLDQ25,
	author       = {Yufei Yang and
	Song Xiao and
	Lixiang Li and
	Wenqian Dong and
	Jiahui Qu},
	title        = {Do You Steal My Model? Signature Diffusion Embedded Dual-Verification
	Watermarking for Protecting Intellectual Property of Hyperspectral
	Image Classification Models},
	booktitle    = {Proceedings of the Thirty-Fourth International Joint Conference on
	Artificial Intelligence, {IJCAI} 2025, Montreal, Canada, August 16-22,
	2025},
	pages        = {2233--2241},
	publisher    = {ijcai.org},
	year         = {2025},
	url          = {https://doi.org/10.24963/ijcai.2025/249},
	doi          = {10.24963/IJCAI.2025/249},
	bibsource    = {dblp computer science bibliography, https://dblp.org}
}

@inproceedings{WangGZ0H0T25,
	author       = {Zilan Wang and
	Junfeng Guo and
	Jiacheng Zhu and
	Yiming Li and
	Heng Huang and
	Muhao Chen and
	Zhengzhong Tu},
	title        = {SleeperMark: Towards Robust Watermark against Fine-Tuning Text-to-image
	Diffusion Models},
	booktitle    = {{IEEE/CVF} Conference on Computer Vision and Pattern Recognition,
	{CVPR} 2025, Nashville, TN, USA, June 11-15, 2025},
	pages        = {8213--8224},
	publisher    = {Computer Vision Foundation / {IEEE}},
	year         = {2025},
	url          = {https://openaccess.thecvf.com/content/CVPR2025/html/Wang\_SleeperMark\_Towards\_Robust\_Watermark\_against\_Fine-Tuning\_Text-to-image\_Diffusion\_Models\_CVPR\_2025\_paper.html},
	doi          = {10.1109/CVPR52734.2025.00769},
	bibsource    = {dblp computer science bibliography, https://dblp.org}
}

@inproceedings{LeiG0Z025,
	author       = {Liangqi Lei and
	Keke Gai and
	Jing Yu and
	Liehuang Zhu and
	Qi Wu},
	title        = {Secure and Efficient Watermarking for Latent Diffusion Models in Model
	Distribution Scenarios},
	booktitle    = {Proceedings of the Thirty-Fourth International Joint Conference on
	Artificial Intelligence, {IJCAI} 2025, Montreal, Canada, August 16-22,
	2025},
	pages        = {7473--7481},
	publisher    = {ijcai.org},
	year         = {2025},
	url          = {https://doi.org/10.24963/ijcai.2025/831},
	doi          = {10.24963/IJCAI.2025/831},
	bibsource    = {dblp computer science bibliography, https://dblp.org}
}

@inproceedings{LiuCQ00P25,
	author       = {Gaozhi Liu and
	Silu Cao and
	Zhenxing Qian and
	Xinpeng Zhang and
	Sheng Li and
	Wanli Peng},
	title        = {Watermarking One for All: {A} Robust Watermarking Scheme Against Partial
	Image Theft},
	booktitle    = {{IEEE/CVF} Conference on Computer Vision and Pattern Recognition,
	{CVPR} 2025, Nashville, TN, USA, June 11-15, 2025},
	pages        = {8225--8234},
	publisher    = {Computer Vision Foundation / {IEEE}},
	year         = {2025},
	url          = {https://openaccess.thecvf.com/content/CVPR2025/html/Liu\_Watermarking\_One\_for\_All\_A\_Robust\_Watermarking\_Scheme\_Against\_Partial\_CVPR\_2025\_paper.html},
	doi          = {10.1109/CVPR52734.2025.00770},
	bibsource    = {dblp computer science bibliography, https://dblp.org}
}

@article{XiaoZHXW24,
	author       = {Xiangli Xiao and
	Yushu Zhang and
	Zhongyun Hua and
	Zhihua Xia and
	Jian Weng},
	title        = {Client-Side Embedding of Screen-Shooting Resilient Image Watermarking},
	journal      = {{IEEE} Trans. Inf. Forensics Secur.},
	volume       = {19},
	pages        = {5357--5372},
	year         = {2024},
	url          = {https://doi.org/10.1109/TIFS.2024.3397043},
	doi          = {10.1109/TIFS.2024.3397043},
	bibsource    = {dblp computer science bibliography, https://dblp.org}
}

@inproceedings{GuoLHGZCWW24,
	author       = {Yiyang Guo and
	Ruizhe Li and
	Mude Hui and
	Hanzhong Guo and
	Chen Zhang and
	Chuangjian Cai and
	Le Wan and
	Shangfei Wang},
	editor       = {Amir Globersons and
	Lester Mackey and
	Danielle Belgrave and
	Angela Fan and
	Ulrich Paquet and
	Jakub M. Tomczak and
	Cheng Zhang},
	title        = {FreqMark: Invisible Image Watermarking via Frequency Based Optimization
	in Latent Space},
	booktitle    = {Advances in Neural Information Processing Systems 38: Annual Conference
	on Neural Information Processing Systems 2024, NeurIPS 2024, Vancouver,
	BC, Canada, December 10 - 15, 2024},
	year         = {2024},
	url          = {http://papers.nips.cc/paper\_files/paper/2024/hash/cbc4912b67d57e3932f56f3fa99faab3-Abstract-Conference.html},
	bibsource    = {dblp computer science bibliography, https://dblp.org}
}

@inproceedings{WenKGG23,
	author       = {Yuxin Wen and
	John Kirchenbauer and
	Jonas Geiping and
	Tom Goldstein},
	editor       = {Alice Oh and
	Tristan Naumann and
	Amir Globerson and
	Kate Saenko and
	Moritz Hardt and
	Sergey Levine},
	title        = {Tree-Rings Watermarks: Invisible Fingerprints for Diffusion Images},
	booktitle    = {Advances in Neural Information Processing Systems 36: Annual Conference
	on Neural Information Processing Systems 2023, NeurIPS 2023, New Orleans,
	LA, USA, December 10 - 16, 2023},
	year         = {2023},
	url          = {http://papers.nips.cc/paper\_files/paper/2023/hash/b54d1757c190ba20dbc4f9e4a2f54149-Abstract-Conference.html},
	bibsource    = {dblp computer science bibliography, https://dblp.org}
}
}

\end{document}